\documentclass[journal=jpclcd,manuscript=letter,layout=twocolumn]{achemso}

\usepackage{mathtools,amssymb,graphicx}
\usepackage[plainpages=false,pdfpagelabels,colorlinks=true,linkcolor=red,urlcolor=blue,citecolor=blue,pdftitle={Title},pdfauthor={},pdfdisplaydoctitle=true,pdfduplex=DuplexFlipLongEdge]{hyperref}
\usepackage{soul} % to use \st for strikeout 
\usepackage[dvipsnames,usenames]{color}
\usepackage[normalem]{ulem}
\usepackage{tabularx}
\usepackage{graphicx}
\usepackage{xcolor}
\usepackage{bbold, verbatim} % to generate identity 1
\usepackage{float}
\graphicspath{ {./figures/} }
\usepackage{tabularx}
\usepackage{rotating}
\usepackage{braket}
\usepackage{multirow}
\usepackage{ulem}
\usepackage{amsmath}

\makeatletter
\def\Hline{
\noalign{\ifnum0=`}\fi\hrule \@height 1pt \futurelet
\reserved@a\@xhline}
\makeatother

\author{Kousuke Nakano}
\email{kousuke_1123@icloud.com}
\affiliation[NIMS]
{Center for Basic Research on Materials (CBRM), National Institute for Materials Science (NIMS), 1-2-1 Sengen, Tsukuba, Ibaraki 305-0047, Japan}

\author{Benjamin X. Shi}
\affiliation[Flatiron]{Initiative for Computational Catalysis, Flatiron Institute, 160 5th Avenue, New York, NY 10010, USA}

\author{Dario Alf{\`e}}
\affiliation[UNiNa]{Dipartimento di Fisica Ettore Pancini, Universit\`a di Napoli Federico II, Monte S. Angelo, I-80126 Napoli, Italy}
\alsoaffiliation[UCL]{Department of Earth Sciences, University College London, Gower Street, London WC1E 6BT, United Kingdom}
\alsoaffiliation[TYC]{Thomas Young Centre and London Centre for Nanotechnology, 17-19 Gordon Street, London WC1H 0AH, United Kingdom}

\author{Andrea Zen}
\email{andrea.zen@unina.it}
\affiliation[UNiNa]{Dipartimento di Fisica Ettore Pancini, Universit\`a di Napoli Federico II, Monte S. Angelo, I-80126 Napoli, Italy}
\alsoaffiliation[UCL]{Department of Earth Sciences, University College London, Gower Street, London WC1E 6BT, United Kingdom}

\title{Assessing the impact of nodal surface optimization in fixed-node diffusion Monte Carlo on non-covalent interactions} 

\begin{document}

\begin{tocentry}
\centering
\includegraphics[width=5.0cm]{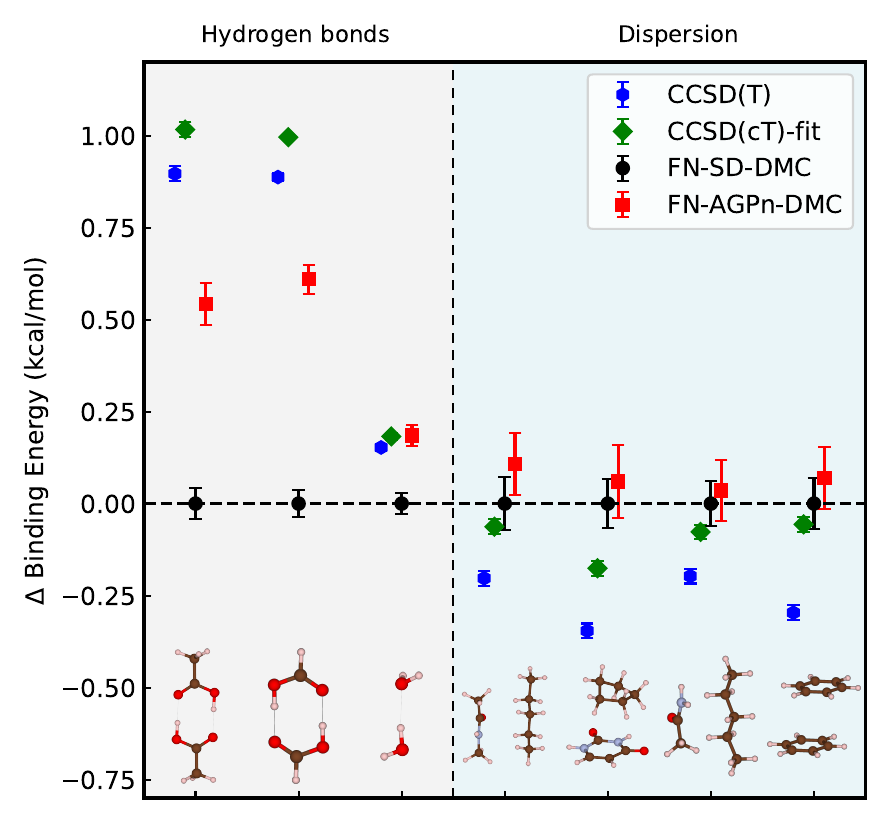}
\label{For Table of Contents Only}
%Some journals require a graphical entry for the Table of Contents. This should be laid out ``print ready'' so that the sizing of the text is correct. Inside the \texttt{tocentry} environment, the font used is Helvetica 8\,pt, as required by \emph{Journal of the American Chemical Society}. The surrounding frame is 9\,cm by 3.5\,cm, which is the maximum permitted for  \emph{Journal of the American Chemical Society} graphical table of content entries. The box will not resize if the content is too big: instead it will overflow the edge of the box. This box and the associated title will always be printed on a separate page at the end of the document.
\end{tocentry}

\begin{abstract}
% This must be no more than 150 words!!
Diffusion quantum Monte Carlo (DMC) and coupled cluster theory [CCSD(T)] are widely-employed benchmark methods for noncovalent interactions (NCIs). However, recent studies have reported notable discrepancies across several hydrogen-bonded and dispersion-dominated systems, raising questions on the accuracy of the approximations underlying each approach. In DMC, the dominant error is expected to stem from the fixed-node approximation, where the nodal surface is typically taken from a single Slater determinant derived from a density functional theory or Hartree–Fock calculation. In this work, we assess the impact of nodal surface optimization on DMC predictions for 12 compounds spanning diverse NCIs, using a recently proposed antisymmetrized geminal power ansatz with natural orbitals. We find improved agreement with CCSD(T) for hydrogen-bonded systems, while having negligible effect for dispersion-dominated systems. These results provide a practical and computationally efficient route to resolving discrepancies in hydrogen-bonded interactions, while offering insight into the remaining differences in dispersion-dominated systems.
\end{abstract}

%%%%%%%%%%%%%%%%%%%%%%%%%%%%%%%%%%%%%%%%%%%%%%%%
%% Introduction
%%%%%%%%%%%%%%%%%%%%%%%%%%%%%%%%%%%%%%%%%%%%%%%%
%\paragraph{{\textit{Introduction}} $-$}
The central task of ab initio electronic structure calculations is to solve the many-body Schrödinger equation. Several methods have been developed to solve this equation, including wavefunction methods (quantum chemistry calculations), density functional theory (DFT), and quantum Monte Carlo (QMC). On the one hand, coupled cluster theory with single, double, and perturbative triple excitations [CCSD(T)]~{\cite{1989_Raghavachari_CCSDT}} has been regarded as a reference method within the wavefunction methods. On the other hand, the Fixed-node Diffusion quantum Monte Carlo with Jastrow Slater Determinant ansatz (FN-SD-DMC)~{\cite{2001FOU}} has also been regarded as a reference method, especially in materials science and condensed matter physics. For small molecular systems, FN-SD-DMC and CCSD(T) typically yield consistent values, e.g., for intermolecular binding energies. However, in 2021, Al-Hamdani et al. reported significant discrepancies between CCSD(T) and FN-SD-DMC calculations when evaluating the binding energies of large molecules bound by van der Waals forces, such as C$_{60}$ buckyball inside a [6]-cycloparaphenyleneacetylene ring (C$_{60}$+5PPA)~{\cite{2021Yasmin_C60}}. 
Fishman et al. reported that such discrepancy is also observed in $\pi$-stacking complexes such as acene and alkadiene dimers~{\cite{2025_Fishman_CCSDT_vdw}}. Furthermore, in 2025, more binding energies were benchmarked on the S66 dataset~{\cite{2025_benjamin_S66}} of medium-sized dimers. It was reported that discrepancies between CCSD(T) and FN-SD-DMC were observed not only for large molecules bound by van der Waals forces but also for some molecules bound by hydrogen bonds.
Since neither CCSD(T) nor FN-SD-DMC provides exact solutions, but rather each employs approximations, this is why no definitive conclusion has been reached regarding which is more accurate. 
%
%%%%%%%%%%%%%%%%%%%%%%%%%%%%%%%%%%%%%%
% Figure
\begin{figure*}[t]
    \centering    \includegraphics[width=1.00\linewidth]{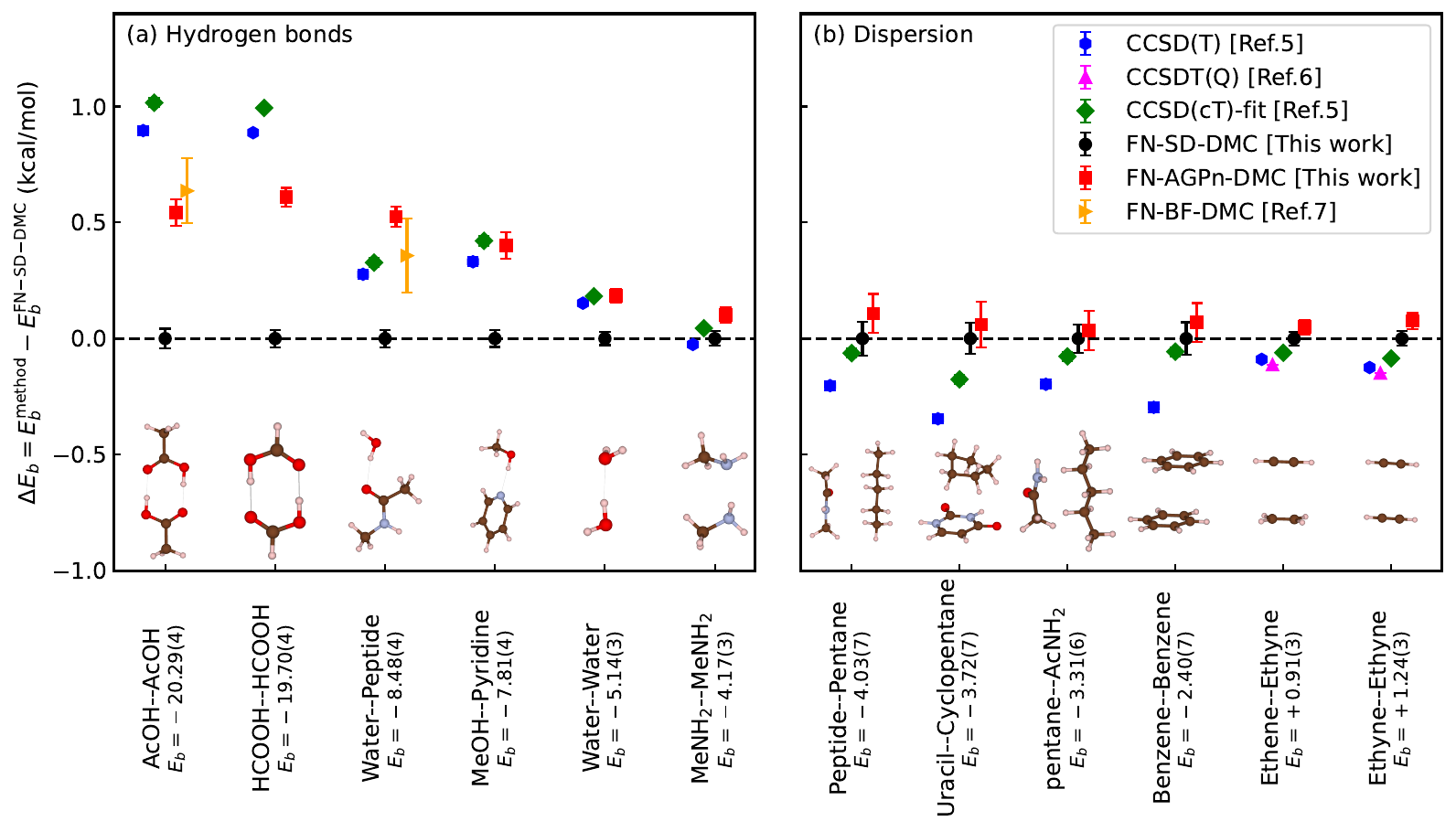}
    \caption[]{The binding energies computed by FN-SD-DMC, FN-AGPn-DMC, FN-BF-DMC, CCSD(T), CCSDT(Q), and CCSD(cT)-fit for (a) hydrogen bonds and (b) Dispersion-dominated bonds. The CCSD(T), CCSDT(Q), CCSD(cT)-fit (except for HCOOH--HCOOH, Ethene--Ethyne, and Ethyne--Ethyne, which were computed in this work using the same setting in Ref.~{\citenum{2025_benjamin_S66}}), and FN-BF-DMC values are taken from Refs.~{\citenum{2025_benjamin_S66}}, {\citenum{2025_Lambie_CCSDTQ}}, {\citenum{2025_benjamin_S66}}, and {\citenum{2025Lambie_arxiv_AcOH}}, respectively. $E_b$ in the $x$ labels refer to the binding energies (kcal/mol) obtained by FN-SD-DMC in this work.}
    \label{fig:E_b_summary}
\end{figure*}
%%%%%%%%%%%%%%%%%%%%%%%%%%%%%%%%%%%%%%

%
\vspace{2mm}
The most uncontrollable approximation in FN-SD-DMC is the fixed-node approximation with a mean-field nodal surface, arising from the so-called Fermion sign problem~{\cite{2001FOU}}. That is, FN-SD-DMC results are determined by the nodal surface provided by a preceding mean-field calculation, such as HF or DFT calculations. In this situation, what can be done from the FN-SD-DMC side is to investigate whether the binding energy obtained by using a {\emph{better}} wavefunction beyond a mean-field nodal surface remains unchanged, or approaches the CCSD(T) value. This allows discussion of whether the discrepancy between CCSD(T) and FN-SD-DMC stems from the mean-field nodal surfaces.

\vspace{2mm}
Many variational ansatzes beyond the single Slater-determinant have been proposed. Examples are multideterminant ansatz~{\cite{1996Filippi_MD, 2005_UMR_MDopt, 2008_Toulouse_MD}}, antisymmetrized geminal power (AGP) ansatz~{\cite{2003_Casula_AGP}}, backflow~{\cite{2006DRU_BF, 2006RIO_BF, 2008Bajdich_PF_BF}}, Pfaffian~{\cite{2006Bajdich_PF, 2008Bajdich_PF_BF, 2020-Genovese-pfaffian}}, and more recently, neural-network-based ansatzes~{\cite{2020-hermann-paulinet, 2020-Pfau-Ferminet}}. They have been applied to binding energy calculations of various molecules~{\cite{2007-Gurtubay-water-BF, 2007_Sorella_adaptive_SR, 2015-Azadi-Benzene-BF, 2019NAK-Na2-AGP, 2020-Genovese-pfaffian, 2023-REN-NN-binding-energy, 2025Lambie_arxiv_AcOH}}.
A pioneering work in the context of the present study is the FN-DMC calculation with the backflow ansatz (denoted as FN-BF-DMC in this study) for AcOH--AcOH and water-peptide by Lambie et al.~{\cite{2025Lambie_arxiv_AcOH}} The study revealed that the consistency between CCSD(T) and FN-BF-DMC is much better than that between CCSD(T) and FN-SD-DMC; thus, the reported discrepancy between CCSD(T) and FN-SD-DMC is attributed to the mean-field nodal surface used in FN-SD-DMC.
However, in any ansatz, the drawback is the inevitable trade-off between flexibility and computational cost. More flexible ansatzes yield lower variational energies but it requires higher computational cost. 
This prevents one from studying diverse compounds at a reasonable computational cost. In fact, Lambie et al.~{\cite{2025Lambie_arxiv_AcOH}} stated that their FN-BF-DMC calculations used 2.9 million and 1.5 million core-hours for AcOH--AcOH and water-peptide on an AMD EPYC cluster machine, respectively, while their FN-SD-DMC counterparts cost 0.42 and 0.27 million core-hours, respectively. This can make it challenging to go towards larger compounds.

\vspace{2mm}
We recently developed a DMC workflow along this line based on AGP ansatz with natural orbitals (denoted as FN-AGPn-DMC in this study)~{\cite{2024_Nakano_FNAGPAS}}, where 
%only the weights in the pairing function in the single determinant 
only a relatively small number of pairing coefficients of the geminal pairing function
are optimized.
%, while the orbitals in the pairing function are frozen to natural orbitals constructed from a beyond HF/DFT calculation. 
This is obtained by writing the pairing function in terms of natural orbitals constructed from a beyond HF/DFT calculation and keeping them frozen in the optimization while optimizing their weights.
Since AGPn is a generalization of a single Slater determinant, FN-AGPn-DMC is variationally better than FN-SD-DMC. Moreover, the computational cost of FN-AGPn-DMC has the same scaling as that of FN-SD-DMC~{\cite{2024_Nakano_FNAGPAS}}. This enables one to study large systems with better nodal surfaces than mean-field ones at an affordable computational cost.

% new paragraph
\vspace{2mm}
This Letter reports DMC binding energy calculations from the FN-AGPn-DMC approach for diverse non-covalent interaction (NCI) compounds, and a comparison with FN-SD-DMC and CCSD(T) evaluations. We selected 12 NCI compounds where discrepancies between FN-SD-DMC and CCSD(T) have been reported. The 12 compounds are taken from the S22~{\cite{2006_PETR_S22}}, S66~{\cite{2011_REZAC_S66}}, and A24~{\cite{2013_REZAC_A24}} datasets, and are divided into two primary categories: hydrogen bonds and dispersion-dominated interactions ($\pi$-stacking, pure London dispersion, and mixed-interactions). Our calculation reveals that, for hydrogen bonds, the FN-AGPn-DMC yields smaller binding energies than the FN-SD-DMC does; thus, provides bonding energies very close to those of CCSD(T). For the dispersion-dominated bonds, even with the better nodal surfaces, FN-AGPn-DMC yields bonding energies consistent with those obtained with FN-SD-DMC. 
Thus, our study shows that the discrepancies between FN-SD-DMC and CCSD(T) in  hydrogen bonds are ultimately attributed to the inaccuracy in the former method, and in particular to the nodal surface of a single Slater determinant.
The same conclusion cannot be straightforwardly generalized to dispersion-dominated bonds, where interaction energies from FN-SD-DMC and FN-AGPn-DMC are in agreement, while the latter improves the wave function and systematically yields lower total energies in all systems.
Thus, identifying the most reliable approach between DMC and CCSD(T) for dispersion-dominated systems remains an open question.

\begin{table*}[t]
\centering
\caption{Binding energies (kcal/mol). The numbers in the parentheses represent 1$\sigma$ error bar. $\Delta E_b^{x} \equiv E_b^{x} - E_b^{\mathrm{FN{\text{-}}SD{\text{-}}DMC}}$, where $x \in \{\mathrm{FN{\text{-}}AGPn{\text{-}}DMC}, \mathrm{CCSD(T)}\}$.}
\label{tab:binding_energy_summary}
\begin{tabular}{l|c c|c|c c}
\hline
System & FN-SD-DMC & FN-AGPn-DMC & CCSD(T) & $\Delta E_b^{\mathrm{FN{\text{-}}AGPn{\text{-}}DMC}}$ & $\Delta E_b^{\mathrm{CCSD(T)}}$ \\
\hline
AcOH--AcOH & -20.29(4) & -19.74(6) & -19.39(2)\cite{2025_benjamin_S66} & 0.54(7) & 0.90(5) \\
HCOOH--HCOOH & -19.70(4) & -19.09(4) & -18.81(1)\cite{2010_Takatani_S22} & 0.61(5) & 0.89(4) \\
Water--Peptide & -8.48(4) & -7.95(4) & -8.20(2)\cite{2025_benjamin_S66} & 0.52(6) & 0.28(4) \\
MeOH--Pyridine & -7.81(4) & -7.41(6) & -7.48(2)\cite{2025_benjamin_S66} & 0.40(7) & 0.33(4) \\
Water--Water & -5.14(3) & -4.96(3) & -4.99(1)\cite{2025_benjamin_S66} & 0.18(4) & 0.15(3) \\
MeNH$_2$--MeNH$_2$ & -4.17(3) & -4.07(3) & -4.20(1)\cite{2025_benjamin_S66} & 0.10(5) & -0.03(3) \\
Peptide--Pentane & -4.03(7) & -3.92(8) & -4.23(2)\cite{2025_benjamin_S66} & 0.1(1) & -0.20(8) \\
Uracil--Cyclopentane & -3.72(7) & -3.7(1) & -4.07(2)\cite{2025_benjamin_S66} & 0.1(1) & -0.35(7) \\
pentane--AcNH$_2$ & -3.31(6) & -3.28(8) & -3.51(2)\cite{2025_benjamin_S66} & 0.0(1) & -0.20(6) \\
Benzene--Benzene & -2.40(7) & -2.33(8) & -2.70(2)\cite{2025_benjamin_S66} & 0.1(1) & -0.30(7) \\
Ethene--Ethyne & 0.91(3) & 0.96(3) & 0.821\cite{2013_Rezac_noncovalent} & 0.05(4) & -0.09(3) \\
Ethyne--Ethyne & 1.24(3) & 1.32(3) & 1.115\cite{2013_Rezac_noncovalent} & 0.08(5) & -0.12(3) \\
\hline
\end{tabular}
\end{table*}

\begin{figure}[t]
    \centering    \includegraphics[width=1.00\linewidth]{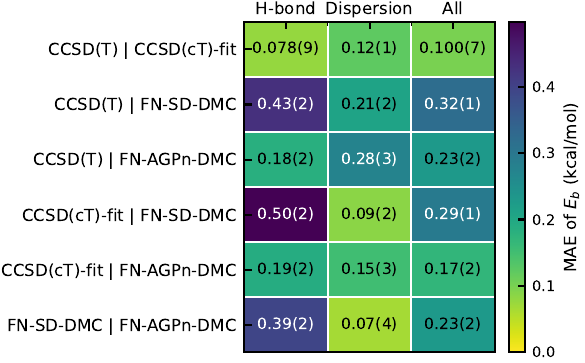}
    \caption[]{Pairwise MAEs of binding energies obtained by CCSD(T), CCSD(cT)-fit, FN-SD-DMC, and FN-AGPn-DMC for hydrogen bonds (H-bond), dispersion-dominated bonds (Dispersion), and all of them (All). The unit is kcal/mol.}
    \label{fig:E_b_MAE}
\end{figure}

\vspace{2mm}
%\paragraph{{\textit{Results and Discussion}} $-$}
%% Result
The key result of this work is summarized in Fig.~{\ref{fig:E_b_summary}} and Table~{\ref{tab:binding_energy_summary}}. Figure {\ref{fig:E_b_summary}} shows the deviation of the binding energy obtained by FN-AGPn-DMC from those obatined by FN-SD-DMC values.
Fig.~{\ref{fig:E_b_summary}} (a) and (b) shows those for the hydrogen and dispersion bonds, respectively. Table~{\ref{tab:binding_energy_summary}} summarizes the binding energy values corresponding to Fig.~{\ref{fig:E_b_summary}}.
For comparison, Fig.~{\ref{fig:E_b_summary}} also plots the deviations of the binding energies obtained by other methods, CCSD(T), CCSDT(Q), CCSD(cT)-fit, and FN-BF-DMC, which will be discussed later.
Figure~{\ref{fig:E_b_MAE}} summarizes the mean absolute errors (MAEs) of binding energies obtained by CCSD(T), CCSD(cT)-fit, FN-SD-DMC, and FN-AGPn-DMC.
On the one hand, it is clear from Fig.~{\ref{fig:E_b_summary}}(a) that, for binding energies for hydrogen bonds, FN-AGPn-DMC weakens the binding energies, resulting closer to those of CCSD(T), specifically for formic acid dimer and acetic acid dimer. While the correction overshoots for Water-Peptide, the consistency between CCSD(T) and FN-AGPn-DMC (0.18(2) kcal/mol in MAD) is better than CCSD(T) and FN-SD-DMC (0.43(2) kcal/mol in MAD), as shown in Fig.~{\ref{fig:E_b_MAE}}. The result implies that, for hydrogen bonds, the reported deviation between FN-SD-DMC and CCSD(T) has arisen from the nodal surfaces obtained from a mean-field approach. The implication agrees with the recent work~{\cite{2025Lambie_arxiv_AcOH}} for AcOH--AcOH and water-peptide.
On the other hand, as shown in Fig.~{\ref{fig:E_b_summary}} (b), for the dispersion-dominated bonds, FN-AGPn-DMC shows negligible effect on binding energy with respect to FN-SD-DMC. That is, for those bonds, the mean-field nodal surfaces are already good enough; thus, the improvement of the nodal surface does not affect binding energies. Our calculation implies that the reported discrepancies between FN-DMC and CCSD(T) for those bonds might not be attributable to a mean-field nodal surface in FN-DMC, suggesting that other factors could be involved. This is discussed later.

\vspace{2mm}
From a practical perspective, it is also important to examine the computational cost of the FN-AGPn-DMC workflows. As described in Ref.~{\citenum{2024_Nakano_FNAGPAS}}, the number of parameters optimized in the FN-AGPn-DMC workflow is limited to the weights associated with the pairing function; thus, it linearly scales with the system size and remains on the order of $10^2$ in practice. For example, even for Uracil–Cyclopentane (dimer), which is the compound with the largest number of valence electrons in this study, the number of optimized parameters is only 183.
Furthermore, AGPn itself is a single determinant of the pairing function. Consequently, the computational cost of FN-AGPn-DMC is not significantly different from that of FN-SD-DMC. For example, for Uracil–Cyclopentane (dimer), the computational cost of FN-AGPn-DMC was 2.2 times that of FN-SD-DMC when achieving the same error bar in total energy (c.f., the computational cost of FN-BF-DMC is approximately 5 $\sim$ 7 times that of FN-SD-DMC~{\cite{2025Lambie_arxiv_AcOH}}).

\vspace{2mm}
Several checks were performed to verify the validity of our calculations; the results are summarized in the Supporting Information. The total energies for all monomers and complexes obtained by FN-AGPn-DMC are lower than those obtained by FN-SD-DMC. The variational principle guarantees that our FN-AGPn-DMC calculations give a better nodal surface than the FN-SD-DMC calculations do. We also confirmed that the underbindings by FN-AGPn-DMC are not due to deterioration in size consistency. CCSD(T) is size-consistent~{\cite{1989_Raghavachari_CCSDT, 1989_Bartlett_CCSDT}}, whereas, in QMC, wavefunction optimization can degrade size-consistency~{\cite{2007_Sorella_adaptive_SR, 2012_Neuscamman_AGP_SC}}, depending on the ansatz choice. Improving the wavefunctions does not necessarily lead to better binding energies because the wavefunctions for the dimer may differ from those for the fragments, leading to artificial under- or over-binding. In this work, we verified that the sum of the energies of the fragments constituting the complex is consistent with the total energy of a far-away complex within the error bars, where the two fragments are sufficiently separated. For all complexes in this study, size consistency is satisfied within the error bars, suggesting that the underbinding is intrinsic.

\vspace{2mm}
%% Discussion
%
Now, let us discuss the origins of the discrepancy between FN-SD-DMC and CCSD(T) binding energies for the non-covalent bonds.
We claim that, for the hydrogen bonds, the reported discrepancies likely due to the employed DFT nodal surface because FN-AGPn-DMC shows significant change in the binding energies and gives closer binding energies with respect to the CCSD(T) values, especially for the hydrogen-bonding systems of Formic Acid dimer and Acetic Acid dimer. 
Our claim is consistent with many recent studies. For instance, Lambie et al. reported the binding energies of Acetic Acid dimer and Water-Peptide~{\cite{2025Lambie_arxiv_AcOH}} by FN-BF-DMC. They revealed that the obtained binding energies are underbound with respect to FN-SD-DMC values. Their FN-BF-DMC binding energies are consistent with our FN-AGPn-DMC ones within the error bars, as shown in Fig.~{\ref{fig:E_b_summary}}. 
Semidalas et al.~{\cite{2025_Semidalas_post_CCSDT}} found that, for hydrogen bonds, CCSD(T) gives consistent binding energies with more accurate CCSDT and CCSDT(Q) calculations. They also argue that the discrepancies reported for the hydrogen-bond systems between FN-SD-DMC and CCSD(T) are very hard to rationalize as remaining flaws of CCSD(T)~{\cite{2025_Semidalas_post_CCSDT}}. This argument is supported by this study.

\vspace{2mm}
The remaining discrepancies ($\sim$ 0.25 kcal/mol) between CCSD(T) and FN-AGPn-DMC can be attributed to the nodal surface error. Indeed, from the DMC side, the only feasible approach is to optimize the nodal surface, as done in this study, because the only uncontrollable error in FN-DMC stems from it. A disclaimer for the present FN-AGPn-DMC calculations is that the ansatz is constrained. Freezing orbitals of the pairing function to the natural orbitals of MP2 and optimizing only the weights is a constraint. For instance, a promising approach beyond this is to relax the constraint with a more flexible ansatz (while ensuring size consistency to enable error cancellation). Specifically, methods such as nodal optimization using neural networks~{\cite{2020-hermann-paulinet, 2020-Pfau-Ferminet}} are likely to be promising, while one should keep in mind that the size-consistency should be checked for variational methods.

\vspace{2mm}
The situation for the dispersion-dominated bonds is different. Our calculations reveal that FN-AGPn-DMC has negligible effect on the binding energy for those bonds compared with FN-SD-DMC; thus, the discrepancy between FN-SD-DMC and CCSD(T) is not explained by the DFT nodal surface, and other sources should contribute to it.
The origin of the discrepancy in those bonds is under intensive debate in the community. One of the pioneering works was done by Sch{\"a}fer et al. reporting changes in binding energies for nine $\pi$-stacking compounds using CCSD(cT)~{\cite{2025_Sch_CCSDCT}}. CCSD(cT) is a CC variant that averts the infrared catastrophe of CCSD(T) by including selected higher-order terms in the triples amplitude approximation without significantly increasing the computational complexity~{\cite{2023_Masion_CCSD_cT}}.
They demonstrated that, for the nine compounds, CCSD(cT) restores excellent agreement for interaction energies with FN-SN-DMC ones.
Based on the way to estimate CCSD(cT) binding energies proposed by Sch{\"a}fer et al. (denoted as CCSD(cT)-fit), Shi et al.~{\cite{2025_benjamin_S66}} reported the CCSD(cT)-fit binding energies for the S66 compounds. The trend is the same as Sch{\"a}fer's work. Indeed, for $\pi$ stacking and other dispersion-dominated compounds, CCSD(cT)-fit gives different binding energies than CCSD(T); thus, CCSD(cT) and CCSD(cT)-fit binding energies are close to FN-SD-DMC ones. The reported and computed CCSD(cT)-fit values are plotted in Fig.~{\ref{fig:E_b_summary}}. Very interestingly, for the tested dispersion bonds, CCSD(cT)-fit gives more consistent binding energies with our FN-SD-DMC and FN-AGPn-DMC ones, implying that only CCSD(T) gives overbindings for those molecules.
However, we notice that, although CCSD(cT) has recently been proposed as an improved alternative to CCSD(T) for non-covalent binding energies, its use as a reference remains under intensive debate, because CCSD(cT) is an approximation to CCSDT, and CCSDT is not always accurate~\cite{2013_Rezac_CCSDT_comparison, 2014_Lucia_CCSDTQ, 2024_Lao_cc_benchmark, 2025_Lambie_CCSDT_large, 2025_Semidalas_post_CCSDT}.
Our work reveals that the discrepancies in the binding energies of the dispersion-dominated interactions might not be explained by the DFT nodal surface used in FN-SD-DMC calculations, but, the origin of the discrepancies remains an open question.

%%%%%%%%%%%%%%%%%%%%%%%%%%%%%%%%%%%%%%
% Figure
%%%%%%%%%%%%%%%%%%%%%%%%%%%%%%%%%%%%%%

%\paragraph{{\textit{Concluding remarks}} $-$}
\vspace{2mm}
In conclusion, we computed the binding energies of 12 compounds with non-covalent bonds, for which their binding energies are reportedly inconsistent between Single, Double, and perturbative Triple excitations [CCSD(T)] and Fixed-Node Diffusion Monte Carlo (FN-DMC) calculations, by the FN-DMC with nodal surfaces given by the Anti-symmetrized Geminal Power ansatz with natural orbitals (AGPn), denoted as FN-AGPn-DMC. The AGPn ansatz yields a better nodal surface than the conventional single Slater Determinant ansatz (SD); thus, FN-AGPn-DMC gives more accurate energies than FN-SD-DMC does, allowing one to discuss whether the discrepancy arises from poor nodal surfaces in FN-DMC. On the one hand, our calculations reveal that, for hydrogen bonds, FN-AGPn-DMC yields underbound binding energies than FN-SD-DMC, leading to greater consistency with CCSD(T). It indicates that the discrepancy previously reported in the binding energy calculations between CCSD(T) and FN-SD-DMC is due to the nodal surface given by a mean-field approach, such as Density Functional Theory (DFT) calculation. On the other hand, our results show that, for dispersion-dominated bonds, FN-AGPn-DMC yields binding energies consistent with those of FN-SD-DMC, thereby not improving the consistency with CCSD(T). It implies that the previously reported discrepancy between CCSD(T) and FN-SD-DMC for dispersion-dominated bonds is not explained by the DFT nodal surface, and the origin of the discrepancies remains an open question. Further investigation will be needed in the community.

\paragraph{{\textit{Computational methods}} $-$}
In this work, we employed our recently developed method for variational optimization of nodal surface with the Fixed-Node Diffusion Monte Carlo (FN-DMC) framework. Typically, in DMC methods, the nodal surface is deterministically given by the single Slater-Determinant constructed from a mean-field calculation, such as HF or DFT, and kept frozen at the DMC level. We denote it FN-SD-DMC in this work. Our developed scheme~{\cite{2024_Nakano_FNAGPAS}} aims to obtain a better wavefunction by variational optimization of the nodal surface at the DMC level. The ansatz choice is the most prominent feature of our scheme. To balance the accuracy and feasibility, we used an Anti-symmetrized geminal power (AGP) wavefunction parametrized with natural orbitals constructed from an MP2 calculation, denoted as AGPn ansatz. The detail is described in Ref.~{\citenum{2024_Nakano_FNAGPAS}}.
In the workflow, the variational optimization of the weights for natural orbitals was performed only for restricted weights, which reduced the number of variational parameters. Regarding how many weights to optimize, we targeted natural orbitals with MP2 calculated occupation ratios $<$ 0.001 for optimization (this is a hyperparameter). 
The AGPn was combined with a conventional Jastrow factor. The Jastrow function in this study is composed of an inhomogeneous one-body, exponential-type two-body, and atomic orbital-based three-body terms~{\cite{2020_Nakano_turborvb}}. The Jastrow factor was optimized only with VMC gradients, and it was held fixed during optimization with (approximate) FN gradients~\cite{2022_McFarland_FNopt}. The weights for the natural orbitals were optimized with the FN gradients. The parameters were optimized using the stochastic reconfiguration method~{\cite{1998_Sorella_SR}} implemented in TurboRVB~{\cite{2020_Nakano_turborvb}} with an adaptive learning rate~{\cite{2007_Sorella_adaptive_SR}}. 

\vspace{2mm}
We took the 12 compound as a benchmark set in this study, taken from the S22, S66, and A24 datasets. They are
the acetic acid dimer (AcOH--AcOH, No.20 in S66),
the formic acid dimer (HCOOH--HCOOH, No.3 in S22),
the dimer of water and peptide (Water--Peptide, No.4 in S66),
the dimer of methanol and pyridine (MeOH--Pyridine, No.19 in S66),
the water dimer (Water--Water, No.1 in S66),
the methylamine dimer (MeNH$_2$--MeNH$_2$, No.10 in S66),
the dimer of peptide and pentane (Peptide--Pentane, No.46 in S66),
the dimer of uracil and cyclopentane (Uracil--Cyclopentane, No.42 in S66),
the dimer of pentane and acetamide (Pentane--AcNH$_2$, No.62 in S66),
the benzene dimer (Benzene--Benzene, No.24 in S66),
the dimer of ethene and ethyne (Ethene--Ethyne, No.22 in A24),
and the ethyne dimer (Ethyne--Ethyne, No.24 in A24).
The compounds are divided into four categories:
hydrogen bonds (Water--Water, HCOOH--HCOOH, AcOH--AcOH, MeNH$_2$-MeNH$_2$, Water--Peptide, and MeOH--Pyridine), 
$\pi$-stacking (Benzene--Benzene), 
pure London dispersion (Uracil--Cyclopentane, Ethene--Ethyne, Ethyne--Ethyne, Peptide--Pentane), and 
mixed-interactions (pentane--AcNH$_2$).
In the main text, we categorize them into two primary groups, hydrogen-bonds and dispersion-dominated compounds ($\pi$-stacking, pure London dispersion, and mixed-interactions).
The target compounds are all dimers composed of two fragments. For each compound, we calculated either three or four configurations: Fragment 1, Fragment 2 (if fragments are not identical), Complex, and Far-away complex (where the two fragments are sufficiently separated; used for size-consistency calculations).

% new patagraph - DMC setup
\vspace{2mm}
For DMC calculations, we employed the correlation consistent effective core potential (ECP)~{\cite{2017_Bennett_ccECP}} and accompanying aug-cc-pVTZ basis sets~{\cite{2017_Bennett_ccECP}}. Our recent study~{\cite{2025_Nakano_BSIE}} has revealed that, even in FN-DMC calculations, the choice of basis function matters for binding energies of the ~0.1 kcal/mol order due to the basis set incompleteness error (BSIE). The aug-cc-pVTZ basis set sufficiently reduces the BSIE~{\cite{2025_Nakano_BSIE}}. For DFT and MP2 calculations for the trial wave function generations, we employed PySCF~{\cite{2018_Sun_pyscf, 2020_Sun_pyscf}}. The LDA-PZ functional was employed for the baseline DFT calculations. We dumped the TREXIO~{\cite{2023_Posenitskiy_TREXIO}} files from PySCF and converted them to the TurboRVB wavefunction format using the converter implemented in TurboGenius~{\cite{2023_Nakano_turbogenius}}. Subsequently, we performed the aforementioned FN-DMC calculations. For DMC calculations, we utilized the lattice regularized diffusion Monte Carlo method (LRDMC)~{\cite{2005_Casula_LRDMC}} implemented in TurboRVB. LRDMC realizes the projection as standard DMC does, and it has been confirmed that controlling all controllable errors yields identical results~{\cite{2025_DellaPia_allQMC, 2025_NAK_LRDMC_load_balanced}}. In LRDMC calculations, the controllable error is the real-space grid, $a$. This error can be eliminated by extrapolating $a \rightarrow 0$. In this study, we adopted values of $a$ = 0.1, 0.15, 0.2, and 0.3 Bohr and fitted the total energy using the quadratic function $E(a) = E_0 + k_1 a^2 + k_2 a^4$. For the total energy, we report the extrapolated value, $E(a \rightarrow 0)$. For binding energies, $E_b = E(\rm{Complex}) - E({\rm{Monomer 1}}) - E(\rm{Monomer 2})$ is computed. We found that the dependence of $E_b$ on $a$ is flat within the error bars for $a \le 0.2$ for all compounds. Therefore, the value at $a = 0.2$ is reported. For size-consistency calculations, $E_s = E(\rm{far{\text{-}}away}) - E({\rm{Monomer 1}}) - E(\rm{Monomer 2})$ is computed. We found that the dependence of $E_s$ on $a$ is also flat within the error bars for $a \le 0.2$ for all compounds. Therefore, the value at $a = 0.2$ is reported. In this study, since ECPs are used, the treatment of the so-called non-local term is crucial to satisfy the variational principle in DMC calculations. Regarding this point, LRDMC corresponds to the implementation of the T-move method in standard DMC~{\cite{2005_Casula_LRDMC, 2006_Casula_Tmove}}, thus ensuring the variational principle. Furthermore, we combined this with the Determinant Locality Approach (DLA)~{\cite{2019_Andrea_DLA}}. This method with the SD ansatz, when starting from the same DFT calculation, yields the same result in the limit $a \rightarrow 0$, i.e., independent of the Jastrow factor parameterization (referred to as the DTM method~{\cite{2025_DellaPia_allQMC}}). All calculations in this study were automated and performed using TurboWorkflows~{\cite{2023_Nakano_turbogenius}}.

\paragraph{{\textit{Acknowledgments}} $-$}
% NIMS computer
K.N. is grateful for computational resources from the Numerical Materials Simulator at National Institute for Materials Science (NIMS).
% FUGAKU
K.N. is grateful for computational resources of the supercomputer Fugaku provided by RIKEN through the HPCI System Research Projects (Project IDs: hp240033 and hp250031).
% K.N. financial support
K.N. acknowledges financial support from MEXT Leading Initiative for Excellent Young Researchers (Grant No.~JPMXS0320220025), from Iketani Science and Technology Foundation (Grant No.~0361248-A), and from JSPS KAKENHI (Grant-in-Aid for Scientific Research(C), Grant No.~25K00216).
D.A. and A.Z. acknowledge support from Leverhulme Trust (Grant No.~RPG-2020-038).
D.A. and A.Z. also acknowledge support from the European Union under the Next generation EU (projects 20222FXZ33 and P2022MC742).
The authors acknowledge the use of the ARCHER UK National Supercomputing Service (https://www.archer2.ac.uk).
% Molecular and Crystal structures (VESTA)
The molecular structures were depicted using VESTA~{\cite{2011MOM}}.
The Flatiron Institute is a division of the Simons Foundation.
We appreciate Prof.~Martin Suhm for his suggestion of the formic acid dimer as an interesting application of our FN-AGPn-DMC approach.

\paragraph{{\textit{Supporting Information}} $-$}
The Supporting Information is available for supplemental data.

\paragraph{{\textit{Code and Data availability}} $-$}
The QMC kernel used in this work, TurboRVB, is available from its GitHub repository, [\url{https://github.com/sissaschool/turborvb}].
TurboRVB is supported by ``Project for advancement of software usability in materials science"~{\cite{Yoshimi2025PASUMS}} in FY2025 by the Institute for Solid State Physics (ISSP) in The University of Tokyo.
The geometries, wavefunctions, pseudo potentials, other input, and output files are available from a GitHub repository, [\url{https://github.com/kousuke-nakano/NCIs-LRDMC-benchmark}].
%{\color{red} [Comment by K.N.] This will be transferred to a public repository, such as ZENODO, during the revision with smaller error bars for $a \ne 0.2$ in SI.}
%
The geometries can also be obtained from the Benchmark Energy and Geometry Database (BEGDB)~{\cite{2008_REZ_BEGDB}}.

%%%%%%%%%%%%%%%%%%%%%%%%%%%%%%%
%\bibliographystyle{apsrev4-1}
\bibliography{./references.bib}

@article{2025_benjamin_S66,
    author = {Shi, Benjamin X. and Della Pia, Flaviano and Al-Hamdani, Yasmine S. and Michaelides, Angelos and Alfè, Dario and Zen, Andrea},
    title = {Systematic discrepancies between reference methods for noncovalent interactions within the S66 dataset},
    journal = {J. Chem. Phys.},
    volume = {162},
    number = {14},
    pages = {144107},
    year = {2025},
    month = {04},
    issn = {0021-9606},
    doi = {10.1063/5.0254021},
    url = {https://doi.org/10.1063/5.0254021},
    eprint = {https://pubs.aip.org/aip/jcp/article-pdf/doi/10.1063/5.0254021/20479764/144107_1_5.0254021.pdf},
}

@article{2010_Takatani_S22,
    author = {Takatani, Tait and Hohenstein, Edward G. and Malagoli, Massimo and Marshall, Michael S. and Sherrill, C. David},
    title = {Basis set consistent revision of the S22 test set of noncovalent interaction energies},
    journal = {J. Chem. Phys.},
    volume = {132},
    number = {14},
    pages = {144104},
    year = {2010},
    month = {04},
    issn = {0021-9606},
    doi = {10.1063/1.3378024},
    url = {https://doi.org/10.1063/1.3378024},
    eprint = {https://pubs.aip.org/aip/jcp/article-pdf/doi/10.1063/1.3378024/16122098/144104_1_online.pdf},
}

@article{2013_Rezac_noncovalent,
author = {Řezáč, Jan and Hobza, Pavel},
title = {Describing Noncovalent Interactions beyond the Common Approximations: How Accurate Is the “Gold Standard,” CCSD(T) at the Complete Basis Set Limit?},
journal = {J. Chem. Theory Comput.},
volume = {9},
number = {5},
pages = {2151-2155},
year = {2013},
doi = {10.1021/ct400057w}
}

@article{2025_Nakano_BSIE,
author = {Nakano, Kousuke and Shi, Benjamin X. and Alfè, Dario and Zen, Andrea},
title = {Basis Set Incompleteness Errors in Fixed-Node Diffusion Monte Carlo Calculations on Noncovalent Interactions},
journal = {J. Chem. Theory Comput.},
volume = {21},
number = {9},
pages = {4426-4434},
year = {2025},
doi = {10.1021/acs.jctc.4c01631},
}

@article{2021Yasmin_C60,
   author = {Yasmine S. Al-Hamdani and Péter R. Nagy and Andrea Zen and Dennis Barton and Mihály Kállay and Jan Gerit Brandenburg and Alexandre Tkatchenko},
   doi = {10.1038/s41467-021-24119-3},
   issn = {20411723},
   issue = {1},
   journal = {Nat. Commun.},
   month = {12},
   pages={3927},
   publisher = {Nature Research},
   title = {Interactions between large molecules pose a puzzle for reference quantum mechanical methods},
   volume = {12},
   year = {2021},
}

@article{2025_Fishman_CCSDT_vdw,
author = {Fishman, Vladimir and Lesiuk, Michał and Martin, Jan M. L. and Daniel Boese, A.},
title = {Another Angle on Benchmarking Noncovalent Interactions},
journal = {J. Chem. Theory and Comput.},
volume = {21},
number = {5},
pages = {2311-2324},
year = {2025},
doi = {10.1021/acs.jctc.4c01512}
}

@article{2025_Sch_CCSDCT,
  title={Understanding discrepancies in noncovalent interaction energies from wavefunction theories for large molecules},
  author={Sch{\"a}fer, Tobias and Irmler, Andreas and Gallo, Alejandro and Gr{\"u}neis, Andreas},
  journal={Nat. Commun.},
  volume={16},
  number={1},
  pages={9108},
  year={2025},
  publisher={Nature Publishing Group UK London}
}

@misc{2025Lambie_arxiv_AcOH,
      title={Nodal error behind discrepancies between coupled cluster and diffusion Monte Carlo in hydrogen-bonded systems}, 
      author={S. Lambie and P. López-Ríos and D. Kats and Ali Alavi},
      year={2026},
      eprint={2508.17937},
      archivePrefix={arXiv},
      primaryClass={physics.chem-ph},
      url={https://arxiv.org/abs/2508.17937}, 
}

@article{2022_McFarland_FNopt,
  title = {Gradient-descent optimization of fermion nodes in the diffusion Monte Carlo technique},
  author = {McFarland, John and Manousakis, Efstratios},
  journal = {Phys. Rev. A},
  volume = {105},
  issue = {3},
  pages = {032815},
  numpages = {11},
  year = {2022},
  month = {Mar},
  publisher = {American Physical Society},
  doi = {10.1103/PhysRevA.105.032815},
  url = {https://link.aps.org/doi/10.1103/PhysRevA.105.032815}
}

@article{1998_Sorella_SR,
	Author = {Sorella, Sandro},
	Journal = {Phys. Rev. Lett.},
	Number = {20},
	Pages = {4558},
	Publisher = {APS},
	Title = {Green function Monte Carlo with stochastic reconfiguration},
	Volume = {80},
	Year = {1998}
}

@article{2007_Sorella_adaptive_SR,
	Author = {Sorella, Sandro and Casula, Michele and Rocca, Dario},
	Journal = {J. Chem. Phys.},
	Number = {1},
	Pages = {014105},
	Publisher = {AIP},
	Title = {Weak binding between two aromatic rings: Feeling the van der Waals attraction by quantum Monte Carlo methods},
	Volume = {127},
	Year = {2007}
}

@article{2020_Nakano_turborvb,
    author = {Nakano, Kousuke and Attaccalite, Claudio and Barborini, Matteo and Capriotti, Luca and Casula, Michele and Coccia, Emanuele and Dagrada, Mario and Genovese, Claudio and Luo, Ye and Mazzola, Guglielmo and Zen, Andrea and Sorella, Sandro},
    title = "{TurboRVB: A many-body toolkit for ab initio electronic simulations by quantum Monte Carlo}",
    journal = {J. Chem. Phys.},
    volume = {152},
    number = {20},
    pages = {204121},
    year = {2020},
    month = {05},
    issn = {0021-9606},
    doi = {10.1063/5.0005037},
    url = {https://doi.org/10.1063/5.0005037},
    eprint = {https://pubs.aip.org/aip/jcp/article-pdf/doi/10.1063/5.0005037/16745553/204121\_1\_online.pdf}
}

@article{2024_Nakano_FNAGPAS,
author = {Nakano, Kousuke and Sorella, Sandro and Alfè, Dario and Zen, Andrea},
title = {Beyond Single-Reference Fixed-Node Approximation in Ab Initio Diffusion Monte Carlo Using Antisymmetrized Geminal Power Applied to Systems with Hundreds of Electrons},
journal = {J. Chem. Theory and Comput.},
volume = {20},
number = {11},
pages = {4591-4604},
year = {2024},
doi = {10.1021/acs.jctc.4c00139},
}

@article{2003_Casula_AGP,
   author = {Michele Casula and Sandro Sorella},
   doi = {10.1063/1.1604379},
   issn = {00225282},
   issue = {13},
   journal = {J. Chem. Phys.},
   month = {10},
   pages = {6500-6511},
   publisher = {American Institute of Physics Inc.},
   title = {Geminal wave functions with Jastrow correlation: A first application to atoms},
   volume = {119},
   year = {2003},
}

@article{2005_Casula_LRDMC,
   author = {Michele Casula and Claudia Filippi and Sandro Sorella},
   doi = {10.1103/PhysRevLett.95.100201},
   issn = {00319007},
   journal = {Phys. Rev. Lett.},
   title = {Diffusion Monte Carlo method with lattice regularization},
   volume = {95},
   issue = {10},
   pages = {100201},
   numpages = {4},
   year = {2005},
   month = {Sep}
}

@article{2006_Casula_Tmove,
   author = {Michele Casula},
   doi = {10.1103/PhysRevB.74.161102},
   issn = {10980121},
   journal = {Phys. Rev. B},
   title = {Beyond the locality approximation in the standard diffusion Monte Carlo method},
   volume = {74},
   issue = {16},
   pages = {161102},
   numpages = {4},
   year = {2006},
   month = {Oct}
}

@article{2017_Bennett_ccECP,
    author = {Bennett, M. Chandler and Melton, Cody A. and Annaberdiyev, Abdulgani and Wang, Guangming and Shulenburger, Luke and Mitas, Lubos},
    title = "{A new generation of effective core potentials for correlated calculations}",
    journal = {J. Chem. Phys.},
    volume = {147},
    number = {22},
    pages = {224106},
    year = {2017},
    month = {12},
    issn = {0021-9606},
    doi = {10.1063/1.4995643},
    url = {https://doi.org/10.1063/1.4995643}
}

@article{2018_Sun_pyscf,
  title={PySCF: the Python-based simulations of chemistry framework},
  author={Sun, Qiming and Berkelbach, Timothy C and Blunt, Nick S and Booth, George H and Guo, Sheng and Li, Zhendong and Liu, Junzi and McClain, James D and Sayfutyarova, Elvira R and Sharma, Sandeep and others},
  journal={Wiley Interdiscip. Rev. Comput. Mol. Sci.},
  volume={8},
  number={1},
  pages={e1340},
  year={2018},
  publisher={Wiley Online Library},
  doi={10.1002/wcms.1340}
}

@article{2020_Sun_pyscf,
  title={Recent developments in the PySCF program package},
   author = {Qiming Sun and Xing Zhang and Samragni Banerjee and Peng Bao and Marc Barbry and Nick S. Blunt and Nikolay A. Bogdanov and George H. Booth and Jia Chen and Zhi Hao Cui and Janus J. Eriksen and Yang Gao and Sheng Guo and Jan Hermann and Matthew R. Hermes and Kevin Koh and Peter Koval and Susi Lehtola and Zhendong Li and Junzi Liu and Narbe Mardirossian and James D. McClain and Mario Motta and Bastien Mussard and Hung Q. Pham and Artem Pulkin and Wirawan Purwanto and Paul J. Robinson and Enrico Ronca and Elvira R. Sayfutyarova and Maximilian Scheurer and Henry F. Schurkus and James E.T. Smith and Chong Sun and Shi Ning Sun and Shiv Upadhyay and Lucas K. Wagner and Xiao Wang and Alec White and James Daniel Whitfield and Mark J. Williamson and Sebastian Wouters and Jun Yang and Jason M. Yu and Tianyu Zhu and Timothy C. Berkelbach and Sandeep Sharma and Alexander Yu Sokolov and Garnet Kin Lic Chan},
  journal={J. Chem. Phys.},
  volume={153},
  number={2},
  pages={024109},
  year={2020},
  publisher={AIP Publishing LLC},
  doi={10.1063/5.0006074}
}

@article{2023_Posenitskiy_TREXIO,
    author = {Posenitskiy, Evgeny and Chilkuri, Vijay Gopal and Ammar, Abdallah and Hapka, Michał and Pernal, Katarzyna and Shinde, Ravindra and Landinez Borda, Edgar Josué and Filippi, Claudia and Nakano, Kosuke and Kohulák, Otto and Sorella, Sandro and de Oliveira Castro, Pablo and Jalby, William and Ríos, Pablo López and Alavi, Ali and Scemama, Anthony},
    title = "{TREXIO: A file format and library for quantum chemistry}",
    journal = {J. Chem. Phys.},
    volume = {158},
    number = {17},
    year = {2023},
    month = {05},
    doi = {10.1063/5.0148161},
    pages = {174801},
}

@article{2023_Nakano_turbogenius,
    author = {Nakano, Kousuke and Kohulák, Oto and Raghav, Abhishek and Casula, Michele and Sorella, Sandro},
    title = "{TurboGenius: Python suite for high-throughput calculations of ab initio quantum Monte Carlo methods}",
    journal = {J. Chem. Phys.},
    volume = {159},
    number = {22},
    pages = {224801},
    year = {2023},
    month = {12},
    issn = {0021-9606},
    doi = {10.1063/5.0179003}
}

@article{2025_DellaPia_allQMC,
    author = {Della Pia, Flaviano and Shi, Benjamin X. and Al-Hamdani, Yasmine S. and Alfé, Dario and Anderson, Tyler A. and Barborini, Matteo and Benali, Anouar and Casula, Michele and Drummond, Neil D. and Dubecký, Matúš and Filippi, Claudia and Kent, Paul R. C. and Krogel, Jaron T. and López Ríos, Pablo and Lüchow, Arne and Luo, Ye and Michaelides, Angelos and Mitas, Lubos and Nakano, Kousuke and Needs, Richard J. and Per, Manolo C. and Scemama, Anthony and Schultze, Jil and Shinde, Ravindra and Slootman, Emiel and Sorella, Sandro and Tkatchenko, Alexandre and Towler, Mike and Umrigar, C. J. and Wagner, Lucas K. and Wheeler, William A. and Zhou, Haihan and Zen, Andrea},
    title = {Reproducibility of fixed-node diffusion Monte Carlo across diverse community codes: The case of water–methane dimer},
    journal = {J. Chem. Phys.},
    volume = {163},
    number = {10},
    pages = {104110},
    year = {2025},
    month = {09},
    doi = {10.1063/5.0272974},
}

@article{2019_Andrea_DLA,
   author = {Andrea Zen and Jan Gerit Brandenburg and Angelos Michaelides and Dario Alfè},
   doi = {10.1063/1.5119729},
   issn = {00219606},
   issue = {13},
   journal = {J. Chem. Phys.},
   month = {10},
   pmid = {31594339},
   publisher = {American Institute of Physics Inc.},
   title = {A new scheme for fixed node diffusion quantum Monte Carlo with pseudopotentials: Improving reproducibility and reducing the trial-wave-function bias},
   volume = {151},
   number = {13},
   pages = {134105},
   year = {2019}
}

@article{2025_Lambie_CCSDTQ,
    author = {Lambie, S. and Rickert, C. and Usvyat, D. and Alavi, A. and Kats, D.},
    title = {Benchmarking distinguishable cluster methods to platinum standard CCSDT(Q) non-covalent interaction energies in the A24 dataset},
    journal = {J. Chem. Phys.},
    volume = {163},
    number = {11},
    pages = {111101},
    year = {2025},
    month = {09},
    issn = {0021-9606},
    doi = {10.1063/5.0280601},
    url = {https://doi.org/10.1063/5.0280601},
    eprint = {https://pubs.aip.org/aip/jcp/article-pdf/doi/10.1063/5.0280601/20700988/111101_1_5.0280601.pdf},
}

@article{2020-hermann-paulinet,
  title={Deep-neural-network solution of the electronic Schr{\"o}dinger equation},
  author={Hermann, Jan and Sch{\"a}tzle, Zeno and No{\'e}, Frank},
  journal={Nat. Chem.},
  volume={12},
  number={10},
  pages={891--897},
  year={2020},
  publisher={Nature Publishing Group UK London}
}

@article{2020-Pfau-Ferminet,
  title = {Ab initio solution of the many-electron Schr\"odinger equation with deep neural networks},
  author = {Pfau, David and Spencer, James S. and Matthews, Alexander G. D. G. and Foulkes, W. M. C.},
  journal = {Phys. Rev. Res.},
  volume = {2},
  issue = {3},
  pages = {033429},
  numpages = {20},
  year = {2020},
  month = {Sep},
  publisher = {American Physical Society},
  doi = {10.1103/PhysRevResearch.2.033429},
  url = {https://link.aps.org/doi/10.1103/PhysRevResearch.2.033429}
}

@article{2025_Semidalas_post_CCSDT,
title = {Post-CCSD(T) corrections in the S66 noncovalent interactions benchmark},
journal = {Chem. Phys. Lett.},
volume = {863},
pages = {141874},
year = {2025},
issn = {0009-2614},
doi = {https://doi.org/10.1016/j.cplett.2025.141874},
url = {https://www.sciencedirect.com/science/article/pii/S0009261425000144},
author = {Emmanouil Semidalas and A. Daniel Boese and Jan M.L. Martin},
}

@article{2023_Masion_CCSD_cT,
  title = {Averting the Infrared Catastrophe in the Gold Standard of Quantum Chemistry},
  author = {Masios, Nikolaos and Irmler, Andreas and Sch\"afer, Tobias and Gr\"uneis, Andreas},
  journal = {Phys. Rev. Lett.},
  volume = {131},
  issue = {18},
  pages = {186401},
  numpages = {6},
  year = {2023},
  month = {Oct},
  publisher = {American Physical Society},
  doi = {10.1103/PhysRevLett.131.186401},
  url = {https://link.aps.org/doi/10.1103/PhysRevLett.131.186401}
}

@article{2017eCEPP,
    author = {Trail, J. R. and Needs, R. J.},
    title = {Shape and energy consistent pseudopotentials for correlated electron systems},
    journal = {The Journal of Chemical Physics},
    volume = {146},
    number = {20},
    pages = {204107},
    year = {2017},
    month = {05},
    issn = {0021-9606},
    doi = {10.1063/1.4984046},
    url = {https://doi.org/10.1063/1.4984046},
    eprint = {https://pubs.aip.org/aip/jcp/article-pdf/doi/10.1063/1.4984046/13610682/204107_1_online.pdf},
}

@article{2020CASINO,
    author = {Needs, R. J. and Towler, M. D. and Drummond, N. D. and López Ríos, P. and Trail, J. R.},
    title = {Variational and diffusion quantum Monte Carlo calculations with the CASINO code},
    journal = {J. Chem. Phys.},
    volume = {152},
    number = {15},
    pages = {154106},
    year = {2020},
    month = {04},
    issn = {0021-9606},
    doi = {10.1063/1.5144288},
    url = {https://doi.org/10.1063/1.5144288},
    eprint = {https://pubs.aip.org/aip/jcp/article-pdf/doi/10.1063/1.5144288/16722061/154106_1_online.pdf},
}

@article{2006_PETR_S22,
  title={Benchmark database of accurate (MP2 and CCSD (T) complete basis set limit) interaction energies of small model complexes, DNA base pairs, and amino acid pairs},
  author={Jure{\v{c}}ka, Petr and {\v{S}}poner, Ji{\v{r}}{\'\i} and {\v{C}}ern{\`y}, Ji{\v{r}}{\'\i} and Hobza, Pavel},
  journal={Phys. Chem. Chem. Phys.},
  volume={8},
  number={17},
  pages={1985--1993},
  year={2006},
  publisher={Royal Society of Chemistry}
}

@article{2011_REZAC_S66,
author = {\v{R}ez\'a\v{c}, Jan and Riley, Kevin E. and Hobza, Pavel},
title = {S66: A Well-balanced Database of Benchmark Interaction Energies Relevant to Biomolecular Structures},
journal = {J. Chem. Theory Comput.},
volume = {7},
number = {8},
pages = {2427-2438},
year = {2011},
doi = {10.1021/ct2002946},
}

@article{2013_REZAC_A24,
  title={Describing noncovalent interactions beyond the common approximations: how accurate is the “gold standard,” CCSD (T) at the complete basis set limit?},
  author={\v{R}ez\'a\v{c}, Jan and Hobza, Pavel},
  journal={J. Chem. Theory Comput.},
  volume={9},
  number={5},
  pages={2151--2155},
  year={2013},
  publisher={ACS Publications},
  doi={10.1021/ct400057w}
}

@article{2006DRU_BF,
   author = {N. D. Drummond and P. López Ríos and A. Ma and J. R. Trail and G. G. Spink and M. D. Towler and R. J. Needs},
   doi = {10.1063/1.2204600},
   issn = {00219606},
   journal = {J. Chem. Phys.},
   month = {6},
   title = {Quantum Monte Carlo study of the Ne atom and the Ne$^+$ ion},
    volume = {124},
    number = {22},
    pages = {224104},
    year = {2006}
}

@article{2006RIO_BF,
  title = {Inhomogeneous backflow transformations in quantum Monte Carlo calculations},
  author = {L\'opez R\'{\i}os, P. and Ma, A. and Drummond, N. D. and Towler, M. D. and Needs, R. J.},
  journal = {Phys. Rev. E},
  volume = {74},
  issue = {6},
  pages = {066701},
  numpages = {15},
  year = {2006},
  month = {Dec},
  publisher = {American Physical Society},
  doi = {10.1103/PhysRevE.74.066701},
  url = {https://link.aps.org/doi/10.1103/PhysRevE.74.066701}
}

@article{2006Bajdich_PF,
   author = {M. Bajdich and L. Mitas and G. Drobný and L. K. Wagner and K. E. Schmidt},
   doi = {10.1103/PhysRevLett.96.130201},
   issn = {00319007},
   issue = {13},
   journal = {Phys. Rev. Lett.},
   title = {Pfaffian pairing wave functions in electronic-structure quantum Monte Carlo simulations},
   volume = {96},
   year = {2006},
}

@article{2008Bajdich_PF_BF,
  title = {Pfaffian pairing and backflow wavefunctions for electronic structure quantum Monte Carlo methods},
  author = {Bajdich, M. and Mitas, L. and Wagner, L. K. and Schmidt, K. E.},
  journal = {Phys. Rev. B},
  volume = {77},
  issue = {11},
  pages = {115112},
  numpages = {13},
  year = {2008},
  month = {Mar},
  publisher = {American Physical Society},
  doi = {10.1103/PhysRevB.77.115112},
  url = {https://link.aps.org/doi/10.1103/PhysRevB.77.115112}
}

@article{1996Filippi_MD,
    author = {Filippi, Claudia and Umrigar, C. J.},
    title = {Multiconfiguration wave functions for quantum Monte Carlo calculations of first‐row diatomic molecules},
    journal = {J. Chem. Phys.},
    volume = {105},
    number = {1},
    pages = {213-226},
    year = {1996},
    month = {07},
    issn = {0021-9606},
    doi = {10.1063/1.471865}
}

@article{2005_UMR_MDopt,
  title = {Energy and Variance Optimization of Many-Body Wave Functions},
  author = {Umrigar, C. J. and Filippi, Claudia},
  journal = {Phys. Rev. Lett.},
  volume = {94},
  issue = {15},
  pages = {150201},
  numpages = {4},
  year = {2005},
  month = {Apr},
  publisher = {American Physical Society},
  doi = {10.1103/PhysRevLett.94.150201},
  url = {https://link.aps.org/doi/10.1103/PhysRevLett.94.150201}
}

@article{2008_Toulouse_MD,
    author = {Toulouse, Julien and Umrigar, C. J.},
    title = {Full optimization of Jastrow–Slater wave functions with application to the first-row atoms and homonuclear diatomic molecules},
    journal = {J. Chem. Phys.},
    volume = {128},
    number = {17},
    pages = {174101},
    year = {2008},
    month = {05},
    issn = {0021-9606},
    doi = {10.1063/1.2908237}
}

@article{2020-Genovese-pfaffian,
   author = {Claudio Genovese and Tomonori Shirakawa and Kousuke Nakano and Sandro Sorella},
   doi = {10.1021/acs.jctc.0c00165},
   issn = {15499626},
   issue = {10},
   journal = {J. Chem. Theory Comput.},
   month = {10},
   pages = {6114-6131},
   pmid = {32804497},
   publisher = {American Chemical Society},
   title = {General Correlated Geminal Ansatz for Electronic Structure Calculations: Exploiting Pfaffians in Place of Determinants},
   volume = {16},
   year = {2020},
}

@article{2007-Gurtubay-water-BF,
    author = {Gurtubay, I. G. and Needs, R. J.},
    title = {Dissociation energy of the water dimer from quantum Monte Carlo calculations},
    journal = {J. Chem. Phys.},
    volume = {127},
    number = {12},
    pages = {124306},
    year = {2007},
    month = {09},
    issn = {0021-9606},
    doi = {10.1063/1.2770711},
    url = {https://doi.org/10.1063/1.2770711}
}

@article{2015-Azadi-Benzene-BF,
    author = {Azadi, Sam and Cohen, R. E.},
    title = {Chemical accuracy from quantum Monte Carlo for the benzene dimer},
    journal = {J. Chem. Phys.},
    volume = {143},
    number = {10},
    pages = {104301},
    year = {2015},
    month = {09},
    issn = {0021-9606},
    doi = {10.1063/1.4930137}
}

@article{2019NAK-Na2-AGP,
author = {Nakano, Kousuke and Maezono, Ryo and Sorella, Sandro},
doi = {10.1021/acs.jctc.9b00295},
issn = {1549-9618},
journal = {J. Chem. Theory Comput.},
pages = {4044--4055},
publisher = {American Chemical Society},
title = {{All-Electron Quantum Monte Carlo with Jastrow Single Determinant Ansatz: Application to the Sodium Dimer}},
volume = {15},
year = {2019}
}

@article{2023-REN-NN-binding-energy,
  title={Towards the ground state of molecules via diffusion Monte Carlo on neural networks},
  author={Ren, Weiluo and Fu, Weizhong and Wu, Xiaojie and Chen, Ji},
  journal={Nat. Commun.},
  volume={14},
  number={1},
  pages={1860},
  year={2023},
  publisher={Nature Publishing Group UK London}
}

@article{1989_Raghavachari_CCSDT,
title = {A fifth-order perturbation comparison of electron correlation theories},
journal = {Chem. Phys. Lett.},
volume = {157},
number = {6},
pages = {479-483},
year = {1989},
issn = {0009-2614},
doi = {https://doi.org/10.1016/S0009-2614(89)87395-6},
url = {https://www.sciencedirect.com/science/article/pii/S0009261489873956},
author = {Krishnan Raghavachari and Gary W. Trucks and John A. Pople and Martin Head-Gordon}
}

@article{1989_Bartlett_CCSDT,
  title={Coupled-cluster approach to molecular structure and spectra: a step toward predictive quantum chemistry},
  author={Bartlett, Rodney J},
  journal={J. Phys. Chem.},
  volume={93},
  number={5},
  pages={1697--1708},
  year={1989},
  publisher={ACS Publications}
}

@article{2012_Neuscamman_AGP_SC,
  title = {Size Consistency Error in the Antisymmetric Geminal Power Wave Function can be Completely Removed},
  author = {Neuscamman, Eric},
  journal = {Phys. Rev. Lett.},
  volume = {109},
  issue = {20},
  pages = {203001},
  numpages = {5},
  year = {2012},
  month = {Nov},
  publisher = {American Physical Society},
  doi = {10.1103/PhysRevLett.109.203001},
  url = {https://link.aps.org/doi/10.1103/PhysRevLett.109.203001}
}

@article{2011MOM,
  title={VESTA 3 for three-dimensional visualization of crystal, volumetric and morphology data},
  author={Momma, Koichi and Izumi, Fujio},
  journal={J. Appl. Crystallogr.},
  volume={44},
  number={6},
  pages={1272--1276},
  year={2011},
  publisher={International Union of Crystallography}
}

@article{2017_Dubecky_BSIE,
author = {Dubecký, Matúš},
title = {Noncovalent Interactions by Fixed-Node Diffusion Monte Carlo: Convergence of Nodes and Energy Differences vs Gaussian Basis-Set Size},
journal = {J. Chem. Theory Comput.},
volume = {13},
number = {8},
pages = {3626-3635},
year = {2017},
doi = {10.1021/acs.jctc.7b00537},
}

@article{2009_Wager_qwalk,
  title={QWalk: A quantum Monte Carlo program for electronic structure},
  author={Wagner, Lucas K and Bajdich, Michal and Mitas, Lubos},
  journal={J. Comput. Phys.},
  volume={228},
  number={9},
  pages={3390--3404},
  year={2009},
  publisher={Elsevier}
}

@article{2017_Burkatzki_BFD_ECP,
    author = {Burkatzki, M. and Filippi, C. and Dolg, M.},
    title = {Energy-consistent pseudopotentials for quantum Monte Carlo calculations},
    journal = {J. Chem. Phys.},
    volume = {126},
    number = {23},
    pages = {234105},
    year = {2007},
    month = {06},
    doi = {10.1063/1.2741534},
}

@article{2013_Rezac_CCSDT_comparison,
author = {Řezáč, Jan and Šimová, Lucia and Hobza, Pavel},
title = {CCSD[T] Describes Noncovalent Interactions Better than the CCSD(T), CCSD(TQ), and CCSDT Methods},
journal = {J. Chem. Theory Comput.},
volume = {9},
number = {1},
pages = {364-369},
year = {2013},
doi = {10.1021/ct3008777},
}

@article{2014_Lucia_CCSDTQ,
  title={Evaluation of composite schemes for CCSDT(Q) calculations of interaction energies of noncovalent complexes},
  author={Demovi{\v{c}}ov{\'a}, Lucia and Hobza, Pavel and {\v{R}}ez{\'a}{\v{c}}, Jan},
  journal={Phys. Chem. Chem. Phys.},
  volume={16},
  number={36},
  pages={19115--19121},
  year={2014},
  publisher={Royal Society of Chemistry}
}

@article{2025_Lambie_CCSDT_large,
    author = {Lambie, S. and Kats, D. and Usvyat, D. and Alavi, A.},
    title = {On the applicability of CCSD(T) for dispersion interactions in large conjugated systems},
    journal = {J. Chem. Phys.},
    volume = {162},
    number = {11},
    pages = {114112},
    year = {2025},
    month = {03},
    issn = {0021-9606},
    doi = {10.1063/5.0246763},
    url = {https://doi.org/10.1063/5.0246763},
    eprint = {https://pubs.aip.org/aip/jcp/article-pdf/doi/10.1063/5.0246763/20443663/114112_1_5.0246763.pdf},
}

@article{2024_Lao_cc_benchmark,
    author = {Lao, Ka Un},
    title = {Canonical coupled cluster binding benchmark for nanoscale noncovalent complexes at the hundred-atom scale},
    journal = {J. Chem. Phys.},
    volume = {161},
    number = {23},
    pages = {234103},
    year = {2024},
    month = {12},
    issn = {0021-9606},
    doi = {10.1063/5.0242359},
    url = {https://doi.org/10.1063/5.0242359},
    eprint = {https://pubs.aip.org/aip/jcp/article-pdf/doi/10.1063/5.0242359/20300193/234103_1_5.0242359.pdf},
}

@article{2001FOU,
	Author = {Foulkes, WMC and Mitas, L and Needs, RJ and Rajagopal, G},
	Journal = {Rev. Mod. Phys.},
	Number = {1},
	Pages = {33},
	Publisher = {APS},
	Title = {Quantum Monte Carlo simulations of solids},
	Volume = {73},
	Year = {2001}
}

@article{2025_NAK_LRDMC_load_balanced,
    author = {Nakano, Kousuke and Sorella, Sandro and Casula, Michele},
    title = {Load-balanced diffusion Monte Carlo method with lattice regularization},
    journal = {J. Chem. Phys.},
    volume = {163},
    number = {19},
    pages = {194117},
    year = {2025},
    month = {11},
    doi = {10.1063/5.0296986}
}

@article{2008_REZ_BEGDB,
  title={Quantum chemical benchmark energy and geometry database for molecular clusters and complex molecular systems (www. begdb. com): a users manual and examples},
  author={{\v{R}}ez{\'a}{\v{c}}, Jan and Jure{\v{c}}ka, Petr and Riley, Kevin E and {\v{C}}ern{\`y}, Ji{\v{r}}{\'\i} and Valdes, Haydee and Pluh{\'a}{\v{c}}kov{\'a}, Krist{\`y}na and Berka, Karel and {\v{R}}ez{\'a}{\v{c}}, Tom{\'a}{\v{s}} and Pito{\v{n}}{\'a}k, Michal and Vondr{\'a}{\v{s}}ek, J and Hobza, P.},
  journal={Collect. Czech. Chem. Commun.},
  volume={73},
  number={10},
  pages={1261--1270},
  year={2008},
  doi={10.1135/cccc20081261}
}

@article{Yoshimi2025PASUMS,
author = {Yoshimi, Kazuyoshi and Motoyama, Yuichi and Aoyama, Tomohiro and Kawamura, Mitsuhiro and Kawashima, Naoki},
title = {Project for advancement of software usability in materials science},
journal = {Sci. technol. adv. material, Meth.},
volume = {5},
number = {1},
year = {2025},
doi = {10.1080/27660400.2025.2564055}
}
%%%%%%%%%%%%%%%%%%%%%%%%%%%%%%%

\end{document}

\begin{abstract}
This file includes the supplementary information for the paper titled ``Assessing the impact of nodal surface optimization in fixed-node diffusion Monte Carlo on non-covalent interactions".
\end{abstract}

\makeatletter
\def\Hline{
\noalign{\ifnum0=`}\fi\hrule \@height 1pt \futurelet
\reserved@a\@xhline}
\makeatother

\makeatletter
\renewcommand{\refname}{}
\renewcommand*{\citenumfont}[1]{#1}
\renewcommand*{\bibnumfmt}[1]{[#1]}
\makeatother

\setcounter{table}{0}
\setcounter{equation}{0}
\setcounter{figure}{0}
\renewcommand{\thepage}{s\arabic{page}}
\renewcommand{\thetable}{S\Roman{table}}
\renewcommand{\thefigure}{S\arabic{figure}}
\renewcommand{\theequation}{S\arabic{equation}}
%\renewcommand{\thelstlisting}{S-\arabic{lstlisting}}

%\listoffigures
%\listoftables

\bgroup
\hypersetup{linkcolor = black}
\listoffigures
\listoftables
\egroup

\section{Supporting Data}
Table~{\ref{tab:jsd_binding_energy_this_vs_previous}} shows the results of the LRDMC calculations using the frozen DFT orbitals in this study and those of a previous study using DMC with the frozen DFT orbitals.  Although different software and implementations were used in each study, the binding energies agree within the statistical errors. This confirms the validity of our calculation results. 
Furthermore, we thoroughly checked size-consistency. Specifically, we verified whether the sum of the energies of the fragments constituting the complex (i.e., Monomer 1 and Monomer 2) is consistent with the total energy of a far-away complex, where the two fragments are sufficiently (10 \AA\ was employed in this study) separated and treated as a single molecule. This is because failure to check the property could lead to situations where the binding energy is underestimated due to the artifact. Specifically, if the complex is optimized to a lower energy solely just because it has more atoms than the individual fragments, the apparent binding energy might be underestimated. In this study, as shown in Table~{\ref{tab:total_energy_and_derived}, we calculated the total energy of several far-away complexes and confirmed that the total energy is consistent with the sum of two fragments within the statistical errors. We notice that the binding energies and size-consistency checks were computed with $a = 0.2$.
The other important check we have done is the total energy comparison between FN-SD-DMC and FN-AGPn-DMC. Table~{\ref{tab:total_energy_and_derived} lists the total energies ($a \rightarrow 0$) obtained for each fragment and complex. These results show that for all compounds, the AGPn ansatze are sufficiently lower in energy compared to the wave functions with the frozen DFT orbitals. This indicates that FN-AGPn-DMC provides a better nodal surface than FN-SD-DMC (from the variational principle).

\begin{table}[t]
\centering
\caption{Binding energies (kcal/mol) obtained by FN-SD-DMC calculations. The calculation settings (basis-set, effective core potential, XC for trial WF, locality approximation, software package) used in this work, Ref.~\citenum{2025_benjamin_S66}, Ref.~\citenum{2017_Dubecky_BSIE}, and Ref.~\citenum{2025_Nakano_BSIE} are (aug-cc-pVTZ~{\cite{2017_Bennett_ccECP}}, ccECP~{\cite{2017_Bennett_ccECP}}, LDA-PZ, T-move with DLA~{\cite{2019_Andrea_DLA}}, TurboRVB~{\cite{2020_Nakano_turborvb}}), (aug-VTZ~{\cite{2017_Burkatzki_BFD_ECP}}, BFD-ECP~{\cite{2017_Burkatzki_BFD_ECP}}, LDA-PZ, T-move~{\cite{2006_Casula_Tmove}}, QWalk~{\cite{2009_Wager_qwalk}}), (plane-wave, eCEPP~{\cite{2017eCEPP}}, LDA-PZ, DLA, CASINO~{\cite{2020CASINO}}), and (aug-cc-pV6Z~{\cite{2017_Bennett_ccECP}}, ccECP~{\cite{2017_Bennett_ccECP}}, LDA-PZ, T-move with DLA~{\cite{2019_Andrea_DLA}}, TurboRVB~{\cite{2020_Nakano_turborvb}}), respectively. The numbers in parentheses represent 1$\sigma$ error bar.}
\label{tab:jsd_binding_energy_this_vs_previous}
\begin{tabular}{l|c|c}
\hline
System & This work & Previous works \\
\hline
AcOH--AcOH & -20.29(4) & -20.30(8)\cite{2025_benjamin_S66} \\
HCOOH--HCOOH & -19.70(4) & -19.6(1)\cite{2017_Dubecky_BSIE} \\
Water--Peptide & -8.48(4) & -8.58(7)\cite{2025_benjamin_S66} \\
MeOH--Pyridine & -7.81(4) & -7.89(7)\cite{2025_benjamin_S66} \\
Water--Water & -5.14(3) & -5.16(3)\cite{2025_benjamin_S66} \\
MeNH$_2$--MeNH$_2$ & -4.17(3) & -4.20(6)\cite{2025_benjamin_S66} \\
Peptide--Pentane & -4.03(7) & -3.82(8)\cite{2025_benjamin_S66} \\
Uracil--Cyclopentane & -3.72(7) & -3.59(9)\cite{2025_benjamin_S66} \\
pentane--AcNH$_2$ & -3.31(6) & -3.08(7)\cite{2025_benjamin_S66} \\
Benzene--Benzene & -2.40(7) & -2.33(7)\cite{2025_benjamin_S66} \\
Ethene--Ethyne & 0.91(3) & 1.04(9)\cite{2025_Nakano_BSIE} \\
Ethyne--Ethyne & 1.24(3) & 1.32(8)\cite{2025_Nakano_BSIE} \\
\hline
\end{tabular}
\end{table}

\clearpage
\begin{sidewaystable*}[t]
\centering
\caption{Total energies (Ha), binding energy [$E_b = E_{\mathrm{Complex}} - (E_{\mathrm{Monomer\,1}} + E_{\mathrm{Monomer\,2}})$] (kcal/mol), and size-consistency check [$E_s = E_{\mathrm{Far\mbox{-}away}} - (E_{\mathrm{Monomer\,1}} + E_{\mathrm{Monomer\,2}})$] (kcal/mol). The numbers in parentheses represent 1$\sigma$ statistical uncertainty. The total energies are extrapolated values ($a \rightarrow 0$), while $E_b$ and $E_c$ are computed with $a = 0.2$.}
\label{tab:total_energy_and_derived}
\begin{tabular}{l|l|cccc|cc}
\hline
System & Ansatz & $E_{\mathrm{Monomer\,1}}$ & $E_{\mathrm{Monomer\,2}}$ & $E_{\mathrm{Complex}}$ & $E_{\mathrm{Far\mbox{-}away}}$ & $E_b$ & $E_s$ \\
\hline
\multirow{2}{*}{AcOH--AcOH} & FN-SD-DMC   & -45.82054(7) & -- & -91.6736(2) & -91.6413(2) & -20.29(4) & -0.05(4) \\
                               & FN-AGPn-DMC & -45.82494(7) & -- & -91.6815(2) & -91.6502(3) & -19.74(6) & -0.09(6) \\
\multirow{2}{*}{HCOOH--HCOOH} & FN-SD-DMC   & -38.93099(5) & -- & -77.8935(1) & -77.8619(1) & -19.70(4) & -0.02(4) \\
                               & FN-AGPn-DMC & -38.93553(5) & -- & -77.9014(2) & -77.8712(1) & -19.09(4) & -0.11(4) \\
\multirow{2}{*}{Water--Peptide} & FN-SD-DMC   & -17.23565(3) & -47.18284(9) & -64.4320(1) & -64.4186(1) & -8.48(4) & -0.10(4) \\
                               & FN-AGPn-DMC & -17.23793(3) & -47.1864(1) & -64.4372(2) & -64.4247(1) & -7.95(4) & 0.01(4) \\
\multirow{2}{*}{MeOH--Pyridine} & FN-SD-DMC   & -24.09033(3) & -41.25271(8) & -65.3555(2) & -65.3430(2) & -7.81(4) & -0.06(4) \\
                               & FN-AGPn-DMC & -24.09277(3) & -41.2571(1) & -65.3617(2) & -65.3499(2) & -7.41(6) & -0.01(6) \\
\multirow{2}{*}{Water--Water} & FN-SD-DMC   & -17.23577(4) & -17.23581(3) & -34.47970(5) & -34.47148(6) & -5.14(3) & -0.07(3) \\
                               & FN-AGPn-DMC & -17.23802(3) & -17.23816(3) & -34.48406(5) & -34.47620(4) & -4.96(3) & -0.04(3) \\
\multirow{2}{*}{MeNH$_2$--MeNH$_2$} & FN-SD-DMC   & -18.58762(3) & -18.58754(3) & -37.18200(8) & -37.1752(1) & -4.17(3) & -0.05(3) \\
                               & FN-AGPn-DMC & -18.58937(3) & -18.58933(4) & -37.1854(1) & -37.1787(1) & -4.07(3) & 0.02(3) \\
\multirow{2}{*}{Peptide--Pentane} & FN-SD-DMC   & -35.5870(1) & -47.18293(9) & -82.7758(4) & -82.7696(4) & -4.03(7) & 0.07(8) \\
                               & FN-AGPn-DMC & -35.5888(1) & -47.1865(1) & -82.7821(6) & -82.7744(6) & -3.92(8) & -0.08(8) \\
\multirow{2}{*}{Uracil--Cyclopentane} & FN-SD-DMC   & -77.0075(1) & -34.37506(8) & -111.3891(4) & -111.3828(4) & -3.72(7) & -0.1(1) \\
                               & FN-AGPn-DMC & -77.0134(2) & -34.37680(9) & -111.3960(5) & -111.3899(5) & -3.7(1) & -0.1(1) \\
\multirow{2}{*}{pentane--AcNH$_2$} & FN-SD-DMC   & -40.31388(9) & -35.5869(1) & -75.9061(4) & -75.9011(4) & -3.31(6) & -0.15(6) \\
                               & FN-AGPn-DMC & -40.3175(1) & -35.5891(2) & -75.9097(5) & -75.9063(6) & -3.28(8) & 0.07(7) \\
\multirow{2}{*}{Benzene--Benzene} & FN-SD-DMC   & -37.6192(2) & -- & -75.2420(5) & -75.2387(5) & -2.40(7) & -0.06(7) \\
                               & FN-AGPn-DMC & -37.6225(3) & -- & -75.2488(6) & -75.2460(6) & -2.33(8) & -0.07(8) \\
\multirow{2}{*}{Ethene--Ethyne} & FN-SD-DMC   & -13.71158(4) & -12.45640(3) & -26.16638(5) & -26.16796(6) & 0.91(3) & -0.04(3) \\
                               & FN-AGPn-DMC & -13.71382(4) & -12.45954(3) & -26.17167(7) & -26.17328(6) & 0.96(3) & -0.01(3) \\
\multirow{2}{*}{Ethyne--Ethyne} & FN-SD-DMC   & -12.45636(4) & -- & -24.91077(5) & -24.91270(5) & 1.24(3) & 0.01(3) \\
                               & FN-AGPn-DMC & -12.45947(3) & -- & -24.91688(6) & -24.91891(6) & 1.32(3) & 0.02(3) \\
\hline
\end{tabular}
\end{sidewaystable*}
\clearpage

%%%%%%%%%%%%%%%%%%%%%%%%%%%%%%%%%%%%%%
% Figure
\begin{figure*}
    \centering\includegraphics[width=1.00\linewidth]{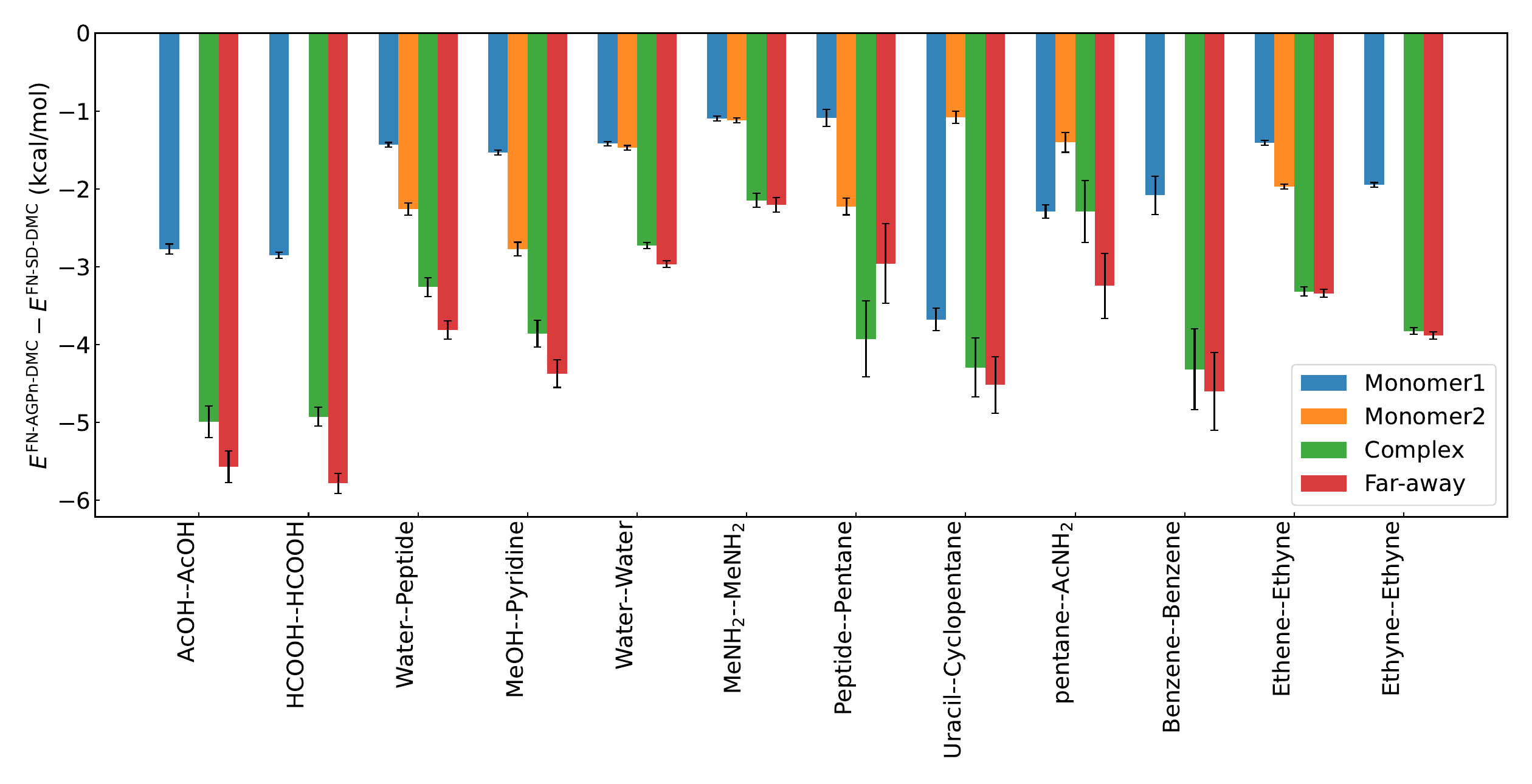}
    \caption{The differences in total energies between FN-SD-DMC and FN-AGPn-DMC for Monomer1, Monomer2 (if it is not identical to Monomer1), Complex, and Far-away complex. The FN-AGPn-DMC yields lower energies than FN-SD-DMC does for all cases, indicating that the obtained nodal surfaces are improved for all cases.}
    \label{fig:E_tot_diff}
\end{figure*}
%%%%%%%%%%%%%%%%%%%%%%%%%%%%%%%%%%%%%%

%%%%%%%%%%%%%%%%%%%%%%%%%%%%%%%%%%%%%%
% Figure
\begin{figure*}
    \centering    \includegraphics[width=1.00\linewidth]{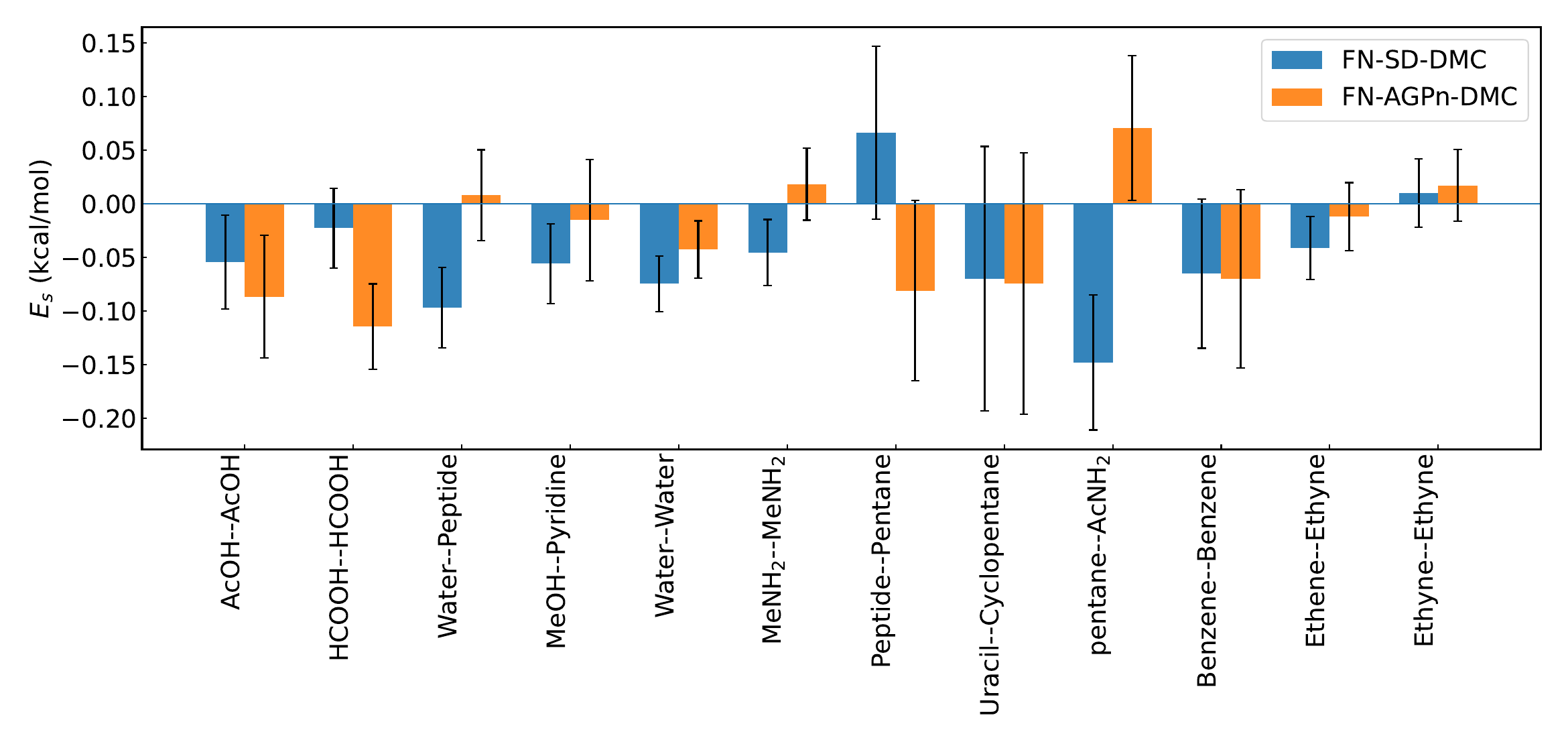}
    \caption{The results for the size-consistency check for FN-SD-DMC and FN-AGPn-DMC. $E_s = E_{\mathrm{Far\mbox{-}away}} - (E_{\mathrm{Monomer\,1}} + E_{\mathrm{Monomer\,2}})$] (kcal/mol). The numbers in the parenthesis represent 1$\sigma$ error bar.}
    \label{fig:E_s}
\end{figure*}
%%%%%%%%%%%%%%%%%%%%%%%%%%%%%%%%%%%%%%

\clearpage

%%%%%%%%%%%%%%%%%%%%%%%%%%%%%%%
%\bibliographystyle{apsrev4-1}
\bibliography{./references.bib}
%%%%%%%%%%%%%%%%%%%%%%%%%%%%%%%